\documentclass[12pt,letterpaper]{article}
\usepackage[utf8]{inputenc}
\usepackage{amsmath}
\usepackage{mathptmx, courier,textcomp}
\usepackage{amsfonts}
\usepackage{amssymb}
\usepackage{graphicx}
\usepackage{cite}
\usepackage{ulem}
\usepackage{caption}
\usepackage{epstopdf}
\usepackage[top=1in,bottom=1in,left=1in,right=1in]{geometry} 
\usepackage{float}
\usepackage{hyperref}
\usepackage{booktabs}
\usepackage[export]{adjustbox}
\usepackage{float}
\usepackage{multirow}
\usepackage{pifont}
\usepackage[ ]{authblk}
\usepackage{chemarrow}
\usepackage{comment}
\usepackage{makecell}
\usepackage{physics}
\UseRawInputEncoding
\hypersetup{
    colorlinks=true,
    linkcolor=blue,
    filecolor=magenta,      
    urlcolor=blue,
    pdftitle={Overleaf Example},
    pdfpagemode=FullScreen,
}

\begin{document}

%
%
\title{Re-Engineering Hematite: Synergistic Co-Doping Routes to Efficient Solar Water Splitting}

%
%

%
%
\author{Abdul Ahad Mamun}
\author[1,*]{Muhammad Anisuzzaman Talukder}
\affil{\small{Department of Electrical and Electronic Engineering\\

Bangladesh University of Engineering and Technology\\

Dhaka 1205, Bangladesh\\}}
\affil[ ]{\small{\it{$^*$anis@eee.buet.ac.bd}}}
%
%

\date{}
\maketitle

\newcommand{\OO}{O$_2$~}
\newcommand{\HH}{H$_2$~}
\sloppy
%
%
\begin{abstract}

Solar-driven water electrolysis requires high-performance photoelectrodes that exhibit excellent photoabsorption, superior charge transport, and optimized thermal management. In this work, we conducted a first-principles investigation to explore optimized doping conditions for hematite ($\alpha$-Fe$_2$O$_3$) by incorporating boron (B), yttrium (Y), and niobium (Nb) mono-dopants, as well as (B, Y) and (B, Nb) co-dopants. To identify the optimal dopant elements and concentrations, we evaluated electronic charge transport, thermal properties, and magnetic susceptibility over a temperature ($T$) range of 300 to 900 K and doping densities ($N$) from $10^{19}$ to $10^{21}$ cm$^{-3}$. The B-doped, (B, Y)-doped, and (B, Nb)-doped $\alpha$-Fe$_2$O$_3$ photoelectrodes showed significantly reduced band gap energy ($E_g$) relative to $\alpha$-Fe$_2$O$_3$. In comparison, Y and Nb dopants only slightly reduced $E_g$ relative to $\alpha$-Fe$_2$O$_3$. While B doping introduced impurity states near the Fermi level that limited thermoelectric charge transport, $\alpha$-Fe$_2$O$_3$ photoelectrodes doped by other elements exhibited notable improvements, including enhanced visible-light absorption, increased carrier concentration, improved electrical conductivity ($\sigma$), and efficient thermal management. Additionally, these doped photoelectrodes exhibited a remarkable increase in Pauli magnetic susceptibility ($\chi$) by two orders of magnitude compared to pristine $\alpha$-Fe$_2$O$_3$, indicating exciting potential for generating spin-selective polarized currents. Overall, our findings revealed that the co-doping conditions are the most effective for enhancing the performance of $\alpha$-Fe$_2$O$_3$, providing a low-cost and high-efficiency solution for sustainable green hydrogen (H$_2$) generation in photocatalytic water splitting.
\end{abstract}
%
%

%
%

%
%
\section{Introduction}
The continuous rise in carbon emissions, particularly from fossil fuel combustion, has caused severe negative impacts on the climate \cite{hait2025emerging}. Therefore, it is essential to reduce carbon emissions and transition to alternative sustainable energy sources to protect the Earth's climate \cite{li2013photoelectrochemical,mamun2024techno}. Solar-driven hydrogen (H$_2$) is a clean and sustainable energy source with the potential to replace traditional fossil fuels \cite{mamun2024enhancing}. H$_2$ offers various applications, is compatible with storage and transportation, and has an energy content that is more than three times greater than that of gasoline \cite{mamun2025advancing,bhuiyan2025hydrogen}. 

Hematite ($\alpha$-Fe$_2$O$_3$) is emerging as a promising photoelectrode for solar-driven H$_2$ generation through water electrolysis, thanks to its low cost, natural abundance, favorable bandgap energy ($E_g$), tunable oxygen atoms, and high stability across a wide range of electrolyte environments \cite{najaf2021recent,mamun2025advancing}. However, the efficient production of H$_2$ using the $\alpha$-Fe$_2$O$_3$ photoelectrode faces complex challenges arising from its optoelectronic and thermophysical limitations. These inherent physical property challenges significantly hinder its performance in photocatalytic water splitting \cite{franking2013facile,tamirat2016using,kim2023thermal}.

The indirect $E_g$ of $\alpha$-Fe$_2$O$_3$ allows it to absorb light up to a wavelength of $\sim 540$ nm, restricting its ability to effectively utilize a significant portion of the solar spectrum \cite{tamirat2016using}. The lower photoabsorption coefficient near the band edge limits light-harvesting efficiency, consequently decreasing photoinduced current. The strong localization of charge carriers in Fe's 3d-orbitals impedes the separation of electron--hole pairs within the bulk, which leads to a higher activation energy requirement for polaronic charge transport \cite{li2018surface}. Additionally, the short diffusion length ($\sim 2$--$4$ nm) and ultrafast recombination rate significantly reduce effective charge carrier concentration ($n_{\rm eff}$) and electrical conductivity ($\sigma$), resulting in inferior electronic charge transport \cite{gardner1963electrical,mamun2025improved}.   

Combining these electronic and charge transport challenges, $\alpha$-Fe$_2$O$_3$ has a low electronic thermal conductivity (${\kappa}^e$) of $< 0.01$ W/mK, which restricts the efficient dissipation of heat during a sustained and prolonged water-splitting process under solar irradiation \cite{shayganpour2021flexible,ito2010electrical}. This effect leads to localized heating that exacerbates phonon scattering, increases defect formation, and disrupts carrier lifetimes by creating additional non-radiative recombination pathways \cite{kim2023thermal}. Furthermore, it degrades the electronic and chemical integrity of the material, destabilizing its catalytic activity in photoelectrochemical (PEC) water splitting. Additionally, the intrinsically low Pauli magnetic susceptibility ($\chi$) of $\alpha$-Fe$_2$O$_3$ indicates weak spin-selective transport and reduced spin polarization of photoexcited carriers. This phenomenon limits charge separation due to spin-blockade effects, ultimately restricting the generation of spin-polarized photoinduced current \cite{dannegger2023magnetic,gajapathy2024spin}.  

Recently, several strategies have been employed to address the challenges associated with $\alpha$-Fe$_2$O$_3$, including nanostructuring, interfacial methods, strain engineering, and element doping \cite{tofanello2020strategies,park2023recent}. The nanostructuring technique uses vertically aligned nanorods or nanowires to enhance photoabsorption and decouple optical absorption from charge transport pathways, thereby maximizing carrier separation and collection efficiency \cite{chnani2021hematite}. The interfacial method incorporates conductive scaffolds or composites onto $\alpha$-Fe$_2$O$_3$, which enhances charge transfer, facilitates efficient thermal management, and promotes long-term stability \cite{xia2021rational,dai2023interfacial}. However, these techniques do not alter the intrinsic properties of $\alpha$-Fe$_2$O$_3$, such as $E_g$, $n_{\rm eff}$, $\sigma$, ${\kappa}^e$, and $\chi$ \cite{mamun2025advancing,osterloh2013inorganic}. In contrast, strain engineering can modify $E_g$ and affect carrier transport dynamics by introducing lattice distortions. However, this technique encounters several limitations, including rapid strain relaxation, mechanical instability, creation of defects and trap sites, and it requires a complex synthesis method \cite{miao2022strain,dai2019strain}. These approaches primarily focus on improving specific performance metrics, along with facing scalability challenges for large-scale applications due to their inherent complexity and limited overall performance.  

On the other hand, element doping has emerged as a promising technique that is more effective, practical, and commercially scalable \cite{zhang2020gradient,al2023water}. This method significantly enhances electronic structures, optical properties, charge transport dynamics, thermal management, and photocatalytic activity simultaneously, making it a superior enhancement strategy compared to other methods \cite{mamun2025advancing}. A wide variety of dopants can be utilized based on their classification in the periodic table, and the manufacturing processes involved are often low-cost and straightforward. While experimental studies are essential for validating material performance, they often face challenges in isolating and interpreting the effects of specific dopants. The complexity of these experimental procedures can hinder the ability to accurately identify how doping influences the intrinsic properties of materials. 

To address the challenges of trial-and-error processes, inconclusive results, and inefficient fabrication pathways, computational simulations---particularly first-principles studies and Boltzmann transport theory---are emerging as essential tools before fabricating doped $\alpha$-Fe$_2$O$_3$ photoelectrodes \cite{guo2015first,akiyama2017effects,butler2019designing}. These computational methods effectively predict and screen dopants, enabling a comprehensive evaluation of optoelectronic and charge carrier properties without the costs and uncertainties associated with physical synthesis \cite{owais2025investigating}. Consequently, the theoretical discovery of novel dopants and co-dopant combinations in pristine $\alpha$-Fe$_2$O$_3$ provides a powerful approach to tailoring its electronic structure, enhancing charge transport, improving thermoelectric performance, and adjusting magnetic characteristics. 

In previous work, we designed doped $\alpha$-Fe$_2$O$_3$ photoelectrodes that included mono-dopants such as boron (B), yttrium (Y), and niobium (Nb), as well as co-dopants like (B, Y) and (B, Nb) \cite{mamun2025improved}. This design involved investigating thermodynamic phase stability, material characterization, electronic structure, and optical absorption through first-principles studies and quantitative computational analysis. Our current study examined charge transport, thermal behavior, and magnetic susceptibility of both pristine and doped $\alpha$-Fe$_2$O$_3$ photoelectrodes to identify optimal doping conditions and understand the underlying physicochemical mechanisms. We simulated these photoelectrodes using a first-principles study based on density functional theory (DFT) combined with semiclassical Boltzmann transport equations, employing the BoltzTraP package. We determined the effective mass of charge carriers ($m^*$) at the valence band maximum (VBM) and conduction band minimum (CBM), as well as the carrier group velocity ($v_g$) from the electronic band structure. Subsequently, we systematically calculated charge transport features, including the Seebeck coefficient ($S$), electronic figure of merit (ZT$_e$), $n_{\rm eff}$, $\sigma$, ${\kappa}^e$, $\chi$, and power factor (PF). 

With the exception of B mono-doping, the doped $\alpha$-Fe$_2$O$_3$ photoelectrodes exhibited a red shift in light absorption, a significant increase in $n_{\rm eff}$, and markedly enhanced $\sigma$. They also demonstrated effective thermal management and improved $\chi$ compared to pristine $\alpha$-Fe$_2$O$_3$. The increase in $\chi$ indicates the potential to generate spin-driven polarized current, which enhances charge separation and facilitates efficient multi-electron reactions during the water-splitting process. These findings confirm that doped $\alpha$-Fe$_2$O$_3$ is a highly promising, low-cost, and high-performance photoelectrode for efficient and sustainable solar-driven H$_2$ generation.
%
%

%
%
\section{Computational Methodology} 

We conducted first-principles simulations using spin-polarized density functional theory (DFT) through a self-consistent ab initio method implemented in the Quantum Espresso (QE) software \cite{giannozzi2009quantum}. The general gradient approximation (GGA) and the Perdew-Burke-Ernzerhof (PBE) model were utilized to estimate the exchange-correlation functions \cite{giannozzi2009quantum,giannozzi2017advanced}. Core electrons were modeled using the projected augmented wave (PAW) methodology. Valence electrons were defined by Kohn-Sham (KS) single-electron orbitals, which we solved by expanding in a plane-wave basis with a kinetic energy cutoff of 60 Ry \cite{kresse1999ultrasoft}. The cutoff for the charge density kinetic energy was approximately twelve times greater than that for the wavefunction. To achieve convergence, we applied Marzari-Vanderbilt smearing with a value of 0.01 Ry. Brillouin-zone integration was performed using generalized Monckhorst-Pack grids of 11$\times$11$\times$1 for geometry optimization and 12$\times$12$\times$12 for electronic structure calculations, which included $E_g$, band structure, and density of states (DOS) \cite{monkhorst1976special}. 

To account for long-range van der Waals interactions, we incorporated the DFT-D3 method developed by Grimme et al.~\cite{grimme2010consistent}. The GGA-PBE calculation tends to inaccurately predict $E_g$ and electronic structures since it neglects self-interaction Coulombic forces within the d-orbitals of transition metals, particularly in transition metal oxides (TMOs) \cite{filippetti2009practical}. To achieve a more accurate estimation of the self-interacting exchange-correlation functions in the d-orbitals of transition metal atoms, we employed the DFT+U framework developed by Dudarev et al.~\cite{dudarev1998electron}. The effective Hubbard correction term, denoted as $U_{\rm eff}$, was set to 4.30, as symmetrically determined by Mosey et al.~\cite{mosey2008rotationally}.

In this study, we selected the optimized and thermodynamically stable geometric structures for doped $\alpha$-Fe$_2$O$_3$ photoelectrodes, as outlined in Ref.~\citeonline{mamun2025improved}. The designed doping concentration ($N_{\rm dd}$) was set at $4.719 \times 10^{20}$ cm$^{-3}$ for the B, Y, and Nb mono-dopants, and at $9.439 \times 10^{20}$ cm$^{-3}$ for the (B, Y) and (B, Nb) co-dopants. To quantify the number of mobile charge carriers that actively contribute to conductivity and photocurrent generation, we calculated $n_{\rm eff}$ by incorporating the DOS using the following equation \cite{madsen2018boltztrap2}
\begin{equation}
    n_{\rm eff} = \int D(E)f(E,T)[1-f(E,T)]dE, 
\end{equation}
where $E$, $T$, and $D(E)$ represent the chemical potential, temperature, and electronic DOS, respectively. $f(E,T)$ is the Fermi-Dirac distribution function.

In terms of $v_g$ across the Cartesian coordinate directions, the energy-projected conductivity tensors, $\sigma_{{\alpha}\beta}(E)$, can be determined using the eigenvalues at each $k$ point obtained from DFT calculations, as given by \cite{akiyama2017effects}
\begin{equation}
    \sigma_{{\alpha}{\beta}}(E) = \frac{q^2}{N_k {\Delta}E}\sum_{i,k} v_{g,\alpha}(i,k)v_{g,\beta}(i,k) \delta(E-E_{i,k}), 
\end{equation}
where $q$, $N_k$, and $E_{i,k}$ represent the elementary electric charge, the number of $k$ points sampled, and the $i^{\rm th}$ eigenvalue at each $k$ point, respectively. The delta function indicates the eigenvalues of $E_{i,k}$ that correspond to energy values distributed within the interval between $E$ and $E + \Delta E$. The product $v_{g,\alpha}(i,k)v_{g,\beta}(i,k)$ is calculated over all electronic eigenstates to evaluate the transport tensor. By utilizing $\sigma_{\alpha\beta}(E)$ and $f(E,T)$, we calculate $\sigma_{\alpha\beta}(E,T)$, $S_{\alpha\beta}(E,T)$, and $\kappa_{\alpha\beta}^e(E,T)$ as a function of $T$ and $E$ within the rigid band approach using the following equations \cite{akiyama2017effects}
\begin{subequations}
\begin{equation}
    \sigma_{{\alpha}\beta}(E,T) = \frac{1}{\Omega}\int \tau \sigma_{\alpha\beta}(E) \left[- \frac{\partial f(E,T)}{\partial E} \right] dE, 
\end{equation}
\begin{equation}
    v_{g,{\alpha}{\beta}}(E,T) = \frac{1}{qT\Omega}\int \tau \sigma_{\alpha\beta}(E) \left(E-E_{\rm F} \right ) \left[- \frac{\partial f(E,T)}{\partial E} \right] dE, 
\end{equation}
\begin{equation}
    S_{{\alpha}{\beta}}(E,T) = \left[\sigma_{i\alpha}(E,T)\right]^{-1}v_{g,i{\beta}}(E,T), 
\end{equation}
\begin{equation}
    \kappa_{{\alpha}{\beta}}^e(E,T) = \frac{1}{q^2 T\Omega}\int \tau \sigma_{\alpha\beta}(E) \left(E-E_{\rm F} \right )^2 \left[- \frac{\partial f(E,T)}{\partial E} \right] dE - T S_{i\alpha}(E,T)v_{g,i{\beta}}(E,T),
\end{equation}
\end{subequations}
where $\Omega$ and $E_{\rm F}$ represent the volume of a unit cell and the Fermi level, respectively. The parameter $\tau$ denotes the relaxation time of the charge carriers. The electronic specific heat, $C(E,T)$, and $\chi(E,T)$ can be derived from the DOS as given by \cite{madsen2006boltztrap,winter2016temperature}
\begin{subequations}
\begin{equation}
    C(E,T) = \int D(E)(E-E_{\rm F}) \left[\frac{\partial f(E,T)}{\partial E} \right] dE, 
\end{equation}
\begin{equation}
    \chi(E,T) = {\mu_0}{\mu_{B}^2}\int D(E)\left[-\frac{\partial f(E,T)}{\partial E} \right] dE, 
\end{equation}
\end{subequations}
where ${\mu_0}$ and ${\mu_{B}}$ are the vacuum permeability and Bohr magneton, respectively.

The values of $S$, $\sigma$, $\kappa^e$, $C$, and $\chi$ were calculated using the BoltzTraP package within the constant relaxation time (CRT) approximation \cite{madsen2006boltztrap,madsen2018boltztrap2}. These calculations were conducted over a temperature range of $300$ to $900$ K to examine the thermal effects on these properties. To evaluate the impact of doping levels on transport behavior, the doping density ($N$) was systematically varied from $10^{19}$ to $10^{21}$ cm$^{-3}$, which represens practical doping conditions for developing a photoelectrode in PEC water splitting. Estimating ZT$_e$ and PF is critical for understanding a material's thermoelectric behavior, including aspects such as the Seebeck response, carrier mobility, and the capacity to convert thermal energy into electrical potential at different temperatures. These parameters directly represent the relationship between charge carrier transport and thermal dissipation, enabling systematic evaluation and optimization of a material's thermoelectric performance. The values of ZT$_e$ and PF can be calculated using the following equations \cite{snyder2008complex}
\begin{subequations}
\begin{equation}
    {\rm ZT}_e = \frac{S^2 \sigma T}{\kappa^e}, 
\end{equation}
\begin{equation}
    {\rm PF} = \frac{S^2 \sigma}{\tau}. 
\end{equation}
\end{subequations}

The PEC system operates within a moderate temperature range and relies on photo-induced electronic processes. These processes include the absorption of photons, the generation of electron-hole pairs (excitons), charge separation, and the transport of charges toward reactive sites. Unlike thermoelectric systems, which primarily depend on temperature gradients to convert thermal energy into electricity, the performance of PEC systems is mainly determined by the electronic band structure and the dynamics of charge transport in the materials under sunlight illumination \cite{chen2013photoelectrochemical}. Consequently, the effect of lattice vibrations (phonons) on charge transport and energy conversion is minimal \cite{cheng2022structural}. Therefore, phonon thermal conductivity---an important factor in high-temperature thermoelectric devices---is not relevant for this PEC application and is excluded from this study.
%
%

%
%
\section{Results and Discussion}

In this section, we analyzed key electronic characteristics, such as $m^*$, $v_g$, and $n_{\rm eff}$, which provide insights into carrier mobility, conductivity, and recombination. To understand the charge transport behavior in doped $\alpha$-Fe$_2$O$_3$ photoelectrodes, it is essential to investigate the impact of carrier dynamics and thermoelectric properties considering the variations of $T$ and $N$. Additionally, we evaluated charge transport properties, including $S$, $\sigma$, $\kappa^e$, and $\chi$, across different $T$ and $N$. Finally, we interpreted the underlying physicochemical mechanisms that contribute to advancements in these properties.    

\subsection{Structural and Electronic Properties}

Determining the corrected optimized lattice parameters and $E_g$ is essential for accurately predicting the optical properties, thermoelectric charge transport, and photocatalytic efficiency of materials. In our simulations, we found the optimized crystal parameters for pristine $\alpha$-Fe$_2$O$_3$ to be $a = b = 5.147$ {\AA} and $c = 13.873$ {\AA}. These values align well with previously published literature, which reported $a = b = 5.104$ {\AA} and $c = 13.907$ {\AA}, demonstrating the reliability of our computational model \cite{meng2016density}. Additionally, the simulated $E_g$ for pristine $\alpha$-Fe$_2$O$_3$ was 2.30 eV, falling within the experimental range of 2.10 eV to 2.40 eV and is consistent with a previously reported value of 2.25 eV \cite{bak2002photo,sivula2011solar,pan2015ti}.  

In this study, we examined three mono-dopants---Y, Nb, and B---and two co-dopants, (B, Y) and (B, Nb), for $\alpha$-Fe$_2$O$_3$. The optimized lattice parameters for all doped $\alpha$-Fe$_2$O$_3$ photoelectrodes showed slight expansion due to several doping-sensitive factors. These factors include differences in atomic sizes of the doping elements, anharmonic lattice vibrations induced by doping, and electronic deformations that affect the band structure \cite{ahmed2020structure}. The Y-doped and Nb-doped $\alpha$-Fe$_2$O$_3$ exhibited slightly decreased $E_g$ to 2.25 eV and 2.18 eV, respectively. In contrast, the B-doped $\alpha$-Fe$_2$O$_3$ demonstrated a significantly lower $E_g$ of 1.65 eV. However, the incorporation of the B dopant introduced impurity states within the band gap, particularly near the Fermi level, which functioned as recombination or trap sites, ultimately hindering charge transport dynamics. 

Notably, the co-doping combinations of (B, Y) and (B, Nb) effectively eliminated these impurity states, resulting in significantly reduced $E_g$ of 1.58 eV and 1.69 eV, respectively. These enhancements promote excellent photoabsorption in the visible spectrum and improve charge transport by minimizing carrier recombination within the material. Based on the Fermi level positions, pristine and B-doped $\alpha$-Fe$_2$O$_3$ were classified as p-type photoelectrodes suitable for use as photocathodes. In contrast, Y-doped, Nb-doped, (B, Y)-doped, and (B, Nb)-doped $\alpha$-Fe$_2$O$_3$ were categorized as n-type photoelectrodes, making them excellent candidates for photoanodes. Table S1 presents the optimized lattice parameters and the values of $E_g$ for pristine and doped $\alpha$-Fe$_2$O$_3$ photoelectrodes, along with $N_{\rm dd}$.  

The effective mass, $m^*$, for the VBM ($m^*_{\rm VBM}$) and CBM ($m^*_{\rm CBM}$), along with the $v_g$, play a crucial role in predicting carrier mobility and diffusion length in materials. Carrier mobility is inversely related to $m^*$ and directly proportional to the square of $v_g$. For pristine $\alpha$-Fe$_2$O$_3$, the calculated $m^*_{\rm CBM}$ was $+2.269m_e$, which is consistent with previously reported values ranging from $+1.50m_e$ to $+3.0m_e$ \cite{meng2013theoretical,xia2013tuning}. The calculated $v_g$ at the CBM for pristine $\alpha$-Fe$_2$O$_3$ was $2.494 \times 10^4$ ms$^{-1}$. In the case of Y-doped $\alpha$-Fe$_2$O$_3$, the $m^*_{\rm CBM}$ value decreased slightly to $+2.096m_e$, while the $v_g$ value increased to $2.597 \times 10^4$ ms$^{-1}$, resulting in improved carrier mobility and diffusion length. 

In contrast, Nb and (B, Nb) dopants exhibited approximately double the $m^*_{\rm CBM}$ compared to pristine $\alpha$-Fe$_2$O$_3$. This increase is attributed to the heavy weight of Nb, which affects the band edges, reduces orbital hybridization between the dopant and host atoms, and enhances spin-orbit coupling, leading to higher $m^*_{\rm CBM}$. On the other hand, B and (B, Y) dopants showed a slight increase in $m^*_{\rm CBM}$ compared to pristine $\alpha$-Fe$_2$O$_3$, and consequently, the $v_g$ at the CBM decreased with these dopants. The values of $m^*_{\rm VBM}$ for all doped $\alpha$-Fe$_2$O$_3$ photoelectrodes were found to be higher than pristine $\alpha$-Fe$_2$O$_3$. These adverse effects arise from the disruption of hybridization between the O's 2p and Fe's 3d orbitals in the valence band, along with altered Fe--O bond lengths and angles due to the presence of heavy and aliovalent dopants \cite{hankin2014constraints}. Such distortions can lead to a reduced kinetic energy of the band curvature near the VBM, resulting in flatter bands and thereby increasing $m^*_{\rm VBM}$. However, these negative effects of $m^*$ can be mitigated by enhancing $\sigma$ and $n_{\rm eff}$, as discussed in later sections. Table \ref{Table1} shows the values of $m^*_{\rm VBM}$, $m^*_{\rm CBM}$, and $v_g$ at the CBM for both pristine and doped $\alpha$-Fe$_2$O$_3$ photoelectrodes.   

%
%
\begin{table}[htbp]
\centering
\caption{The effective mass of charge carrier ($m^*$) at the valence band maxima (VBM) and the conduction band minima (CBM), as well as carrier group velocity ($v_{g}$) at the CBM for pristine and doped $\alpha$-Fe$_2$O$_3$ photoelectrodes.}
\resizebox{0.57\textwidth}{!}{%
\begin{tabular}{c c c c}
\Xhline{3\arrayrulewidth}
    Name & $v_{g}$ & $m_{\rm VBM}^*$ & $m_{\rm CBM}^*$  \\
      &  (ms$^{-1}$) & ($^a$$m_e$) & ($^a$$m_e$)  \\
    \Xhline{2\arrayrulewidth}
    pristine  $\alpha$-Fe$_2$O$_3$  & $2.494 \times 10^{4}$ & $-1.324$ & $2.269$  \\
    Y-doped $\alpha$-Fe$_2$O$_3$    & $2.597 \times 10^{4}$ & $-2.491$ & $2.096$  \\
    Nb-doped $\alpha$-Fe$_2$O$_3$   & $2.972 \times 10^{3}$ & $-2.146$ & $5.643$  \\
    B-doped $\alpha$-Fe$_2$O$_3$    & $1.414 \times 10^{4}$ & $-8.655$ & $3.167$  \\
    (B, Y)-doped $\alpha$-Fe$_2$O$_3$   &  $1.254 \times 10^{4}$ & $-2.079$ & $3.789$  \\
    (B, Nb)-doped $\alpha$-Fe$_2$O$_3$  &  $7.303 \times 10^{3}$ & $-4.484$ & $5.605$  \\
    \Xhline{3\arrayrulewidth}
    \multicolumn{4}l{{$^a$$m_e$ is the static electron mass.}} 
\end{tabular}}
\label{Table1}
\end{table}
%
%

%
\subsection{Thermoelectric Effect on Charge Carrier}

The thermoelectric parameters, particularly $S$ and the dimensionless ZT$_e$, provide valuable insights into the carrier transport, recombination dynamics, and energy conversion efficiency in photoelectrode materials \cite{wang2018first}. The Seebeck coefficient, $S$, measures the thermoelectric voltage generated per unit temperature gradient and reflects the entropy associated with each charge carrier. It is highly sensitive to the chemical energy-dependent distribution of carriers near the Fermi level. A higher $S$ indicates stronger energy selectivity for carriers in photoelectrode materials, which leads to enhanced photoinduced voltage and improved separation of charge carriers. However, the coefficient $S$ is inversely related to carrier concentration, which can negatively affect $\sigma$. Therefore, a moderate to high $S$ value typically reduces carrier recombination and improves overall charge transport dynamics. 

\begin{figure}[htbp]
    \centering
    \includegraphics[width =0.99\linewidth]{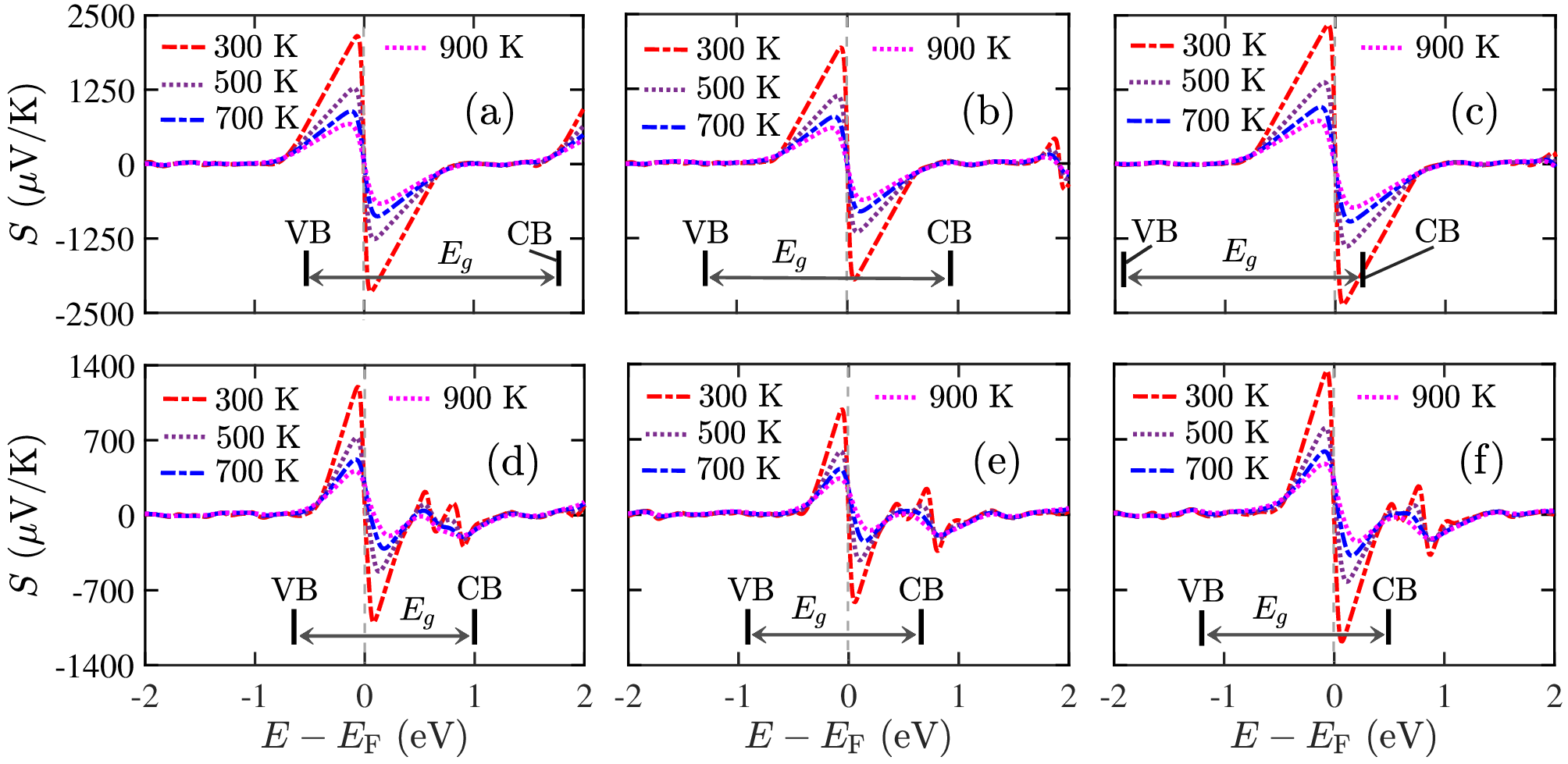}
    \caption{Seebeck coefficient ($S$) as a function of chemical potential ($E$) for (a) pristine $\alpha$-Fe$_2$O$_3$, (b) Y-doped $\alpha$-Fe$_2$O$_3$, (c) Nb-doped $\alpha$-Fe$_2$O$_3$, (d) B-doped $\alpha$-Fe$_2$O$_3$, (e) (B, Y)-doped $\alpha$-Fe$_2$O$_3$, and (f) (B, Nb)-doped $\alpha$-Fe$_2$O$_3$ at different temperatures ($T$) of 300 K, 500 K, 700 K, and 900 K. CB, VB, and $E_g$ represent the conduction band, valence band, and band gap energy, respectively. For all of them, the Fermi level ($E_{\rm F}$) is set to zero energy.}
    \label{Fig1}
\end{figure}

Figure \ref{Fig1} illustrates the $S$ values as a function of $E$, with $T$ varying from 300 to 900 K for both pristine and doped $\alpha$-Fe$_2$O$_3$ photoelectrodes. The maximum $S$ values for pristine, Y-doped, and Nb-doped $\alpha$-Fe$_2$O$_3$ were found to be 2150, 1939, and 2318 {\textmu}V/K at $T=300$ K, respectively, with the corresponding $E$ values distributed within $\pm 0.72$ eV. This result demonstrates superior carrier energy selectivity and a significant enhancement in photoinduced voltage. As $T$ rises, the maximum values of $S$ continuously decline due to several intrinsic factors, including increased carrier excitation, changes in scattering mechanisms, and reduced carrier entropy under degenerate conditions. Notably, at $T=900$ K, the minimum values of $S$ for these photoelectrodes remained above 600 {\textmu}V/K, indicating moderate values and highlighting the favorable thermoelectric effects on charge transport dynamics. 

%
%
%

The B-doped, (B, Y)-doped, and (B, Nb)-doped $\alpha$-Fe$_2$O$_3$ exhibited maximum values of $S$ that were approximately half of pristine $\alpha$-Fe$_2$O$_3$ at $T=300$ K. This notable reduction can be attributed to degenerate doping mechanisms, increased carrier concentrations, and a decrease in charge carrier entropy arising from the shift of the Fermi level toward the conduction band (CB). At $T=900$ K, the minimum value of $S$ remained above $350$ {\textmu}V/K for these photoelectrodes, highlighting their suitability for efficient photovoltage generation in PEC systems, as well as potential applications in thermoelectric devices.

Additionally, the appearance of extra peaks near the CB (positive energy sides) for these photoelectrodes suggests a substantial improvement in chemical energy-dependent carrier transport. These characteristics reflect complex band structures resembling impurity-induced resonant states, which may optimize carrier selectivity and charge separation efficiency. We also investigated the effects of $N$ on $S$ for both pristine and doped $\alpha$-Fe$_2$O$_3$ photoelectrodes (refer to Fig.~S1). The coefficient $S$ displayed negligible variation in doped $\alpha$-Fe$_2$O$_3$ photoelectrodes, while pristine $\alpha$-Fe$_2$O$_3$ exhibited a transition from negative to positive values with increasing $N$. These findings suggest that the doped $\alpha$-Fe$_2$O$_3$ photoelectrodes possess remarkable properties in terms of $S$, paving the way for advancements in carrier transport and photoinduced voltage generation.
%
\begin{figure}[htbp]
    \centering
    \includegraphics[width =1\linewidth]{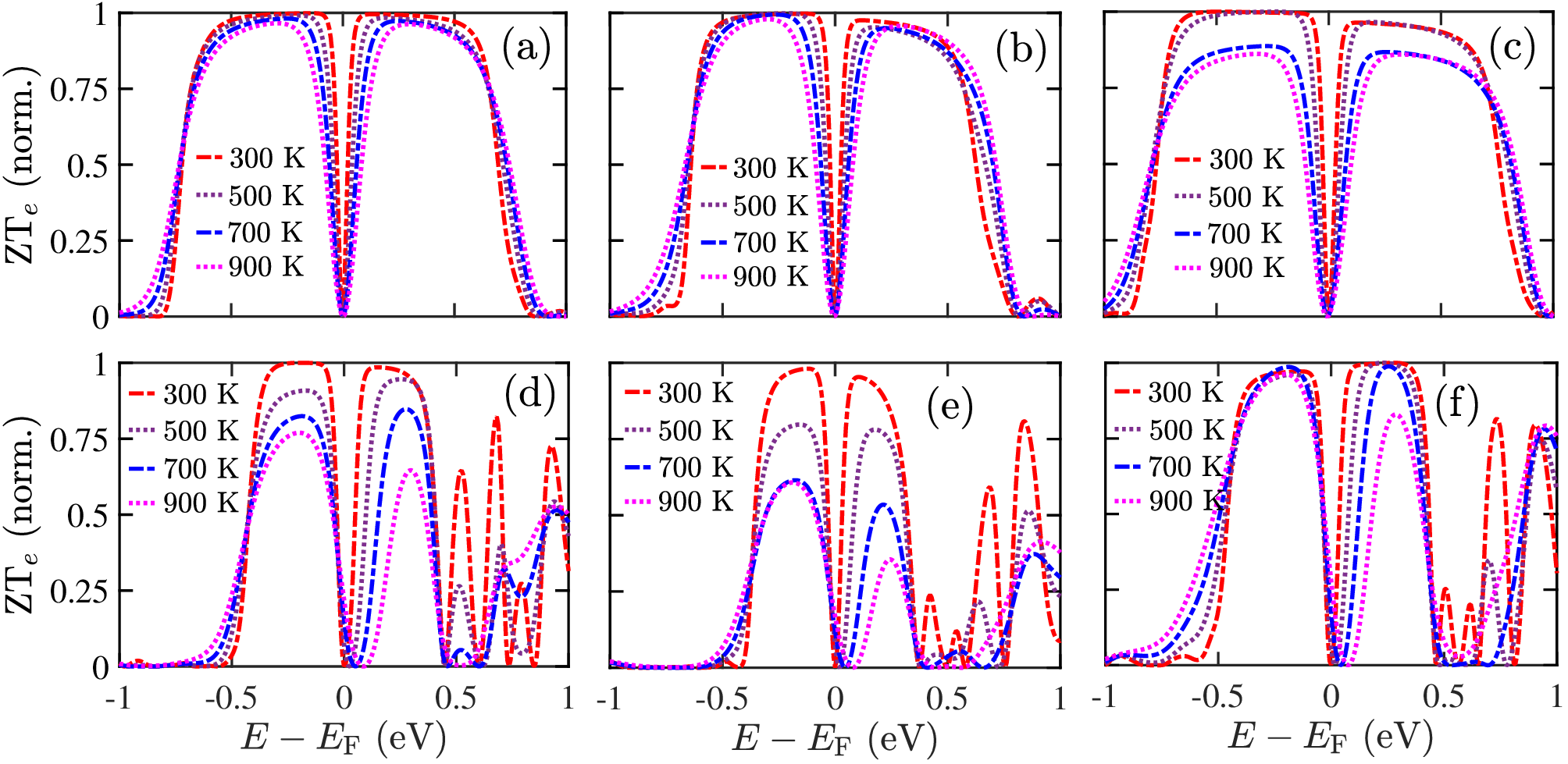}
    \caption{Normalized figure of merit (ZT$_e$) as a function of chemical potential (E) for (a) pristine $\alpha$-Fe$_2$O$_3$, (b) Y-doped $\alpha$-Fe$_2$O$_3$, (c) Nb-doped $\alpha$-Fe$_2$O$_3$, (d) B-doped $\alpha$-Fe$_2$O$_3$, (e) (B, Y)-doped $\alpha$-Fe$_2$O$_3$, and (f) (B, Nb)-doped $\alpha$-Fe$_2$O$_3$ at different temperatures ($T$) of 300 K, 500 K, 700 K, and 900 K. For all of them, the Fermi level ($E_{\rm F}$) is set to zero energy.}
    \label{Fig2}
\end{figure}
%
%

The parameter ZT$_e$ quantifies the energy sensitivity in charge transport within both the valence band (VB) and CB in terms of $\sigma$ and $\kappa^e$. Figure \ref{Fig2} illustrates the normalized ZT$_e$ values for pristine and doped $\alpha$-Fe$_2$O$_3$ photoelectrodes over temperature range of 300 K to 900 K, highlighting the impact of $T$ on carrier selectivity and energy-dependent charge transport. At 300 K, the maximum ZT$_e$ values reported for the different samples were 0.99954 for pristine, 0.99956 for Y-doped, 1.01680 for Nb-doped, 1.01870 for B-doped, 1.00670 for (B, Y)-doped, and 1.0250 for (B, Nb)-doped $\alpha$-Fe$_2$O$_3$ photoelectrodes. These results emphasize the potential application of both pristine and doped $\alpha$-Fe$_2$O$_3$ photoelectrodes in thermoelectric and photoelectrochemical devices.

Pristine and Y-doped $\alpha$-Fe$_2$O$_3$ exhibited ZT$_e$ values close to unity, reflecting a moderate balance between electronic transport and heat dissipation during thermal-to-electric energy conversion. In contrast, B-doping, as well as co-doping with (B, Y) and (B, Nb), resulted in ZT$_e$ values slightly surpassing unity, suggesting a more effective synergy between charge transport and thermal properties. Generally, higher ZT$_e$ values signify enhanced carrier mobility and strong thermoelectric driving forces, which facilitate carrier separation and reduce recombination losses. Additionally, a higher ZT$_e$ correlates with the position of the Fermi level closer to transport-relevant electronic states, thereby enabling shorter carrier pathways and more efficient charge transfer across the photoelectrode--electrolyte interface \cite{xu2021enhanced}. Nonetheless, the overall effectiveness of electronic properties and thermal-to-electrical energy conversion ultimately depends on maximizing $S$, improving $\sigma$, and optimizing carrier concentration.

The Y and Nb dopants exhibited a broader energy distribution with higher ZT$_e$ in both the CB and VB, suggesting improved transport of electrons and holes that is less sensitive to $E$. This insensitivity indicates their stability against the alterations of $E$, which can arise from various factors such as doping levels, scattering mechanisms, and self-generated defects. Conversely, the B, (B, Y), and (B, Nb) dopants displayed a narrower ZT$_e$ profile with multiple sharp peaks on the positive energy side, indicating complex band dispersion and high-energy-sensitive charge carriers in the CB. The impact of $T$ on ZT$_e$ was negligible for pristine and Y-doped $\alpha$-Fe$_2$O$_3$ photoelectrodes, while the ZT$_e$ of the Nb-doped  $\alpha$-Fe$_2$O$_3$ slightly decreased at $T=700$ K and $900$ K. In contrast, the ZT$_e$ values for the B, (B, Y), and (B, Nb) dopants decreased as $T$ increased due to the heightened intrinsic exciton concentrations and reduced $S$. Overall, the Y-doped $\alpha$-Fe$_2$O$_3$ and Nb-doped $\alpha$-Fe$_2$O$_3$ photoelectrodes demonstrated less sensitivity to energy-dependent carriers and exhibited effective thermoelectric properties. However, for the B-doped, (B, Y)-doped, and (B, Nb)-doped $\alpha$-Fe$_2$O$_3$ photoelectrodes, it is essential to carefully optimize the doping levels and material integrity to maintain optimal charge transport across the anticipated operational range of $T$.  

%
\begin{figure}[htbp]
    \centering
    \includegraphics[width =0.83\linewidth]{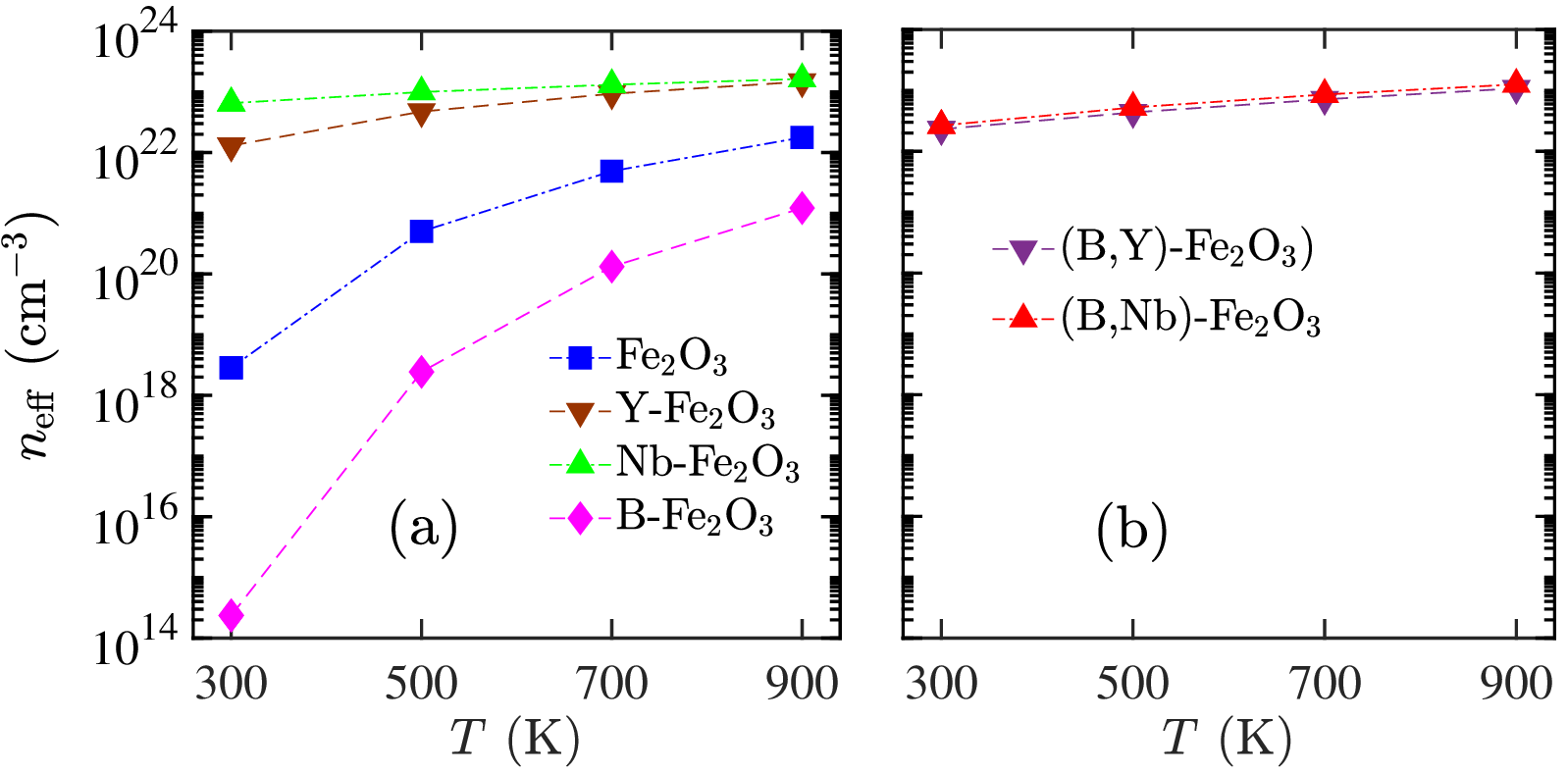}
    \caption{Effective charge carrier concentration ($n_{\rm eff}$) vs. temperature ($T$) for (a) pristine, Y-doped, Nb-doped, and B-doped $\alpha$-Fe$_2$O$_3$ photoelectrodes, and (b) (B, Y)-doped and (B, Nb)-doped $\alpha$-Fe$_2$O$_3$ photoelectrodes.}
    \label{Fig3}
\end{figure}
%
%

%
\subsection{Effective Charge Carrier and Transport Properties}

The analysis of carrier concentration within doped photoelectrode materials is essential for understanding the electronic properties of semiconductors. Dopants introduce additional mobile carriers, which shift the Fermi level and cause band bending at the interface between the photoelectrode and electrolyte. This band bending significantly enhances charge transfer efficiency by creating a built-in electric field. Estimating $n_{\rm eff}$  allows the prediction of the Debye length, the width of the space-charge region, and the electric field strength, all of which influence the drift-diffusion transport of photogenerated carriers. Furthermore, $n_{\rm eff}$ is directly related to $\sigma$, photoinduced voltage, and recombination rates. 

Figure \ref{Fig3} illustrates the $n_{\rm eff}$ values for pristine and doped $\alpha$-Fe$_2$O$_3$ photoelectrodes across a temperature range from 300 to 900 K. The $n_{\rm eff}$ values of doped $\alpha$-Fe$_2$O$_3$ photoelectrodes improved with increasing $T$ due to degenerate carrier behavior driven by thermal excitation. At $T = 300$ K, the measured $n_{\rm eff}$ for pristine $\alpha$-Fe$_2$O$_3$ was $2.843 \times 10^{18}$ cm$^{-3}$, while the values for Y-doped and Nb-doped $\alpha$-Fe$_2$O$_3$ were $1.30 \times 10^{22}$ cm$^{-3}$ and $6.512 \times 10^{22}$ cm$^{-3}$, respectively, indicating significant improvements in the electronic properties of doped photoelectrodes. Additionally, the $n_{\rm eff}$ values for these mono-doped photoelectrodes increased more than nine times at higher $T$ compared to the pristine $\alpha$-Fe$_2$O$_3$ photoelectrode. 

For the co-dopant combinations of (B, Y) and (B, Nb), the $n_{\rm eff}$ values exhibited a trend similar to the Y and Nb mono-dopants with increasing $T$, as shown in Fig.~\ref{Fig3}(b). At $T = 300$ K, the $n_{\rm eff}$ values for the (B, Y) and (B, Nb) co-dopants exceeded those of pristine $\alpha$-Fe$_2$O$_3$ by more than four orders of magnitude. This result indicates improved carrier transport by reducing carrier recombination and lowering internal resistance. It is worth noting that the extremely high $n_{\rm eff}$ values may lead to decreased carrier mobility due to increased scattering of ionized impurities. However, the enhanced $\sigma$ can mitigate these negative effects, as $\sigma$ increases linearly with higher $n_{\rm eff}$. Among all the doping conditions, B-doped $\alpha$-Fe$_2$O$_3$ demonstrated the minimum $n_{\rm eff}$ of $2.362 \times 10^{14}$ cm$^{-3}$ at $T = 300$ K, which is significantly lower than that of pristine $\alpha$-Fe$_2$O$_3$. Notably, Nb dopant contributed a higher number of electrons, resulting in the maximum $n_{\rm eff}$ compared to other doped photoelectrodes in this study. This enhancement underscores the effectiveness of doping in optimizing photoelectrode performance, paving the way for a more efficient photocatalytic water-splitting process. 

%
\begin{figure}[htbp]
    \centering
    \includegraphics[width =1.0\linewidth]{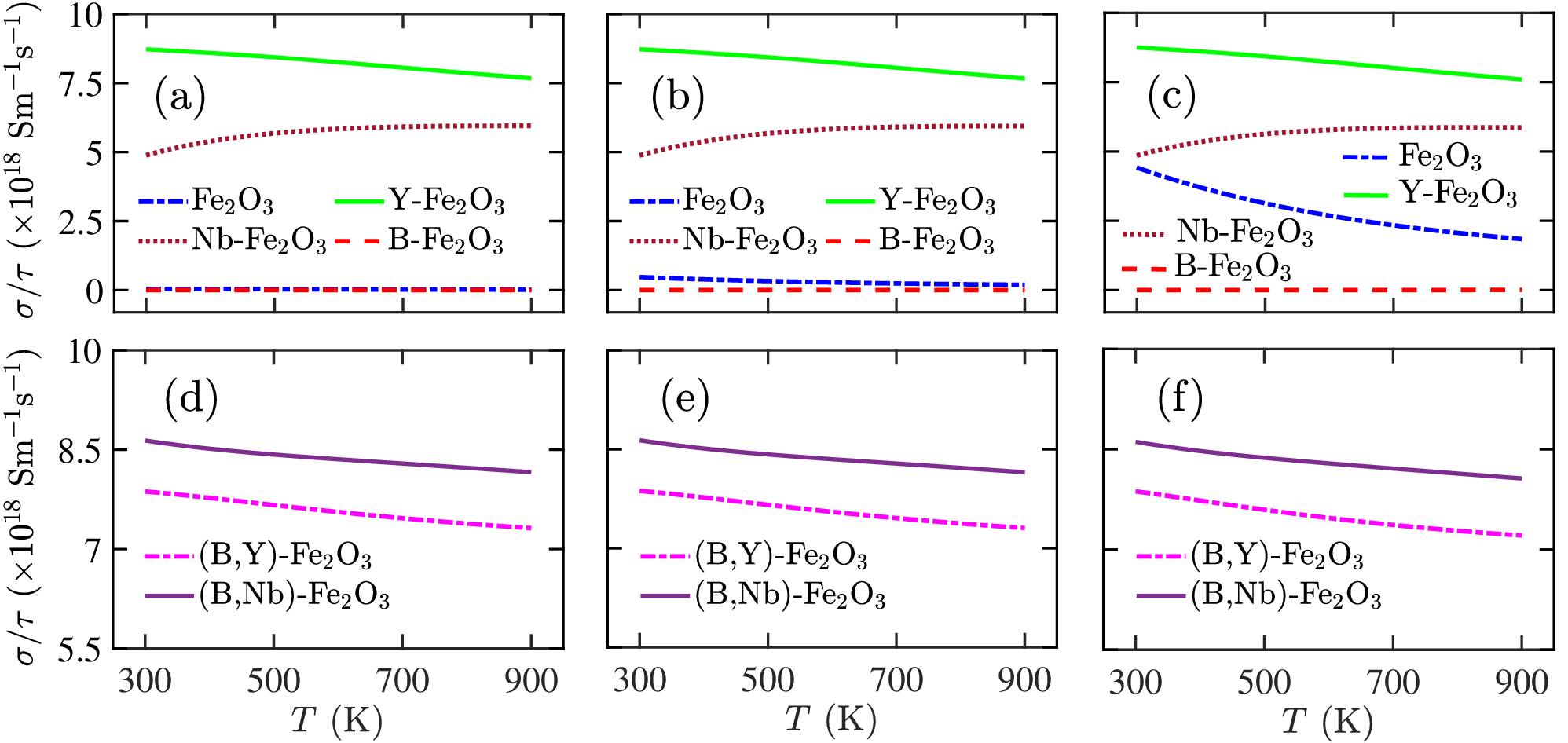}
    \caption{Electrical conductivity divided by relaxation time ($\sigma$/$\tau$) as a function of temperature ($T$) for pristine pristine $\alpha$-Fe$_2$O$_3$, Y-doped $\alpha$-Fe$_2$O$_3$, Nb-doped $\alpha$-Fe$_2$O$_3$, and B-doped $\alpha$-Fe$_2$O$_3$ at various doping densities ($N$) of (a) $10^{19}$ cm$^{-3}$, (b) $10^{20}$ cm$^{-3}$, and (c) $10^{21}$ cm$^{-3}$. Electrical conductivity divided by relaxation time ($\sigma$/$\tau$) as a function of temperature ($T$) for (B,Y)-doped $\alpha$-Fe$_2$O$_3$ and (B,Nb)-doped $\alpha$-Fe$_2$O$_3$ at various doping densities ($N$) of (a) $10^{19}$ cm$^{-3}$, (b) $10^{20}$ cm$^{-3}$, and (c) $10^{21}$ cm$^{-3}$.}
    \label{Fig4}
\end{figure}
%
%

The ratio of $\sigma$ to $\tau$ ($\sigma/\tau$) values for B-doped, Y-doped, and Nb-doped $\alpha$-Fe$_2$O$_3$ showed minimal changes in $N$ across the range of $10^{19}$ to $10^{21}$ cm$^{-3}$, as illustrated in Fig.~\ref{Fig4}(a)--(c). Throughout the range of $T$ from $300$ to $900$ K, the $\sigma/\tau$ value for B doping remained constant due to the presence of impurity bands near the Fermi level. In contrast, the $\sigma/\tau$ values for Y and Nb dopants exhibited decreasing and increasing trends, respectively. These differences can be attributed to the distinct thermal excitation of the trivalent Y and pentavalent Nb doping elements. Additionally, we calculated $\sigma/\tau$ for $\alpha$-Fe$_2$O$_3$ with a monovalent doping element at $N$ ranging from $10^{19}$ to $10^{21}$ cm$^{-3}$, as shown in Fig.~\ref{Fig4}(a)--(c). At $N=10^{21}$ cm$^{-3}$, the $\sigma/\tau$ value for $\alpha$-Fe$_2$O$_3$ was higher than at lower $N$ due to the effects of thermal excitation. Notably, both Y and Nb dopants demonstrated significantly higher $\sigma/\tau$ values compared to pristine $\alpha$-Fe$_2$O$_3$ with a monovalent dopant across all doping scenarios, reinforcing the advantages of the proposed doping strategies.       

Similarly, both (B, Y) and (B, Nb) co-dopants showed significantly higher $\sigma/\tau$ values across a range of $N$ from $10^{19}$ to $10^{21}$ cm$^{-3}$ when compared to pristine $\alpha$-Fe$_2$O$_3$ with a monovalent dopant, as illustrated in Fig.~\ref{Fig4}(d)--(f). The $\sigma/\tau$ values displayed a decreasing trend for these doping profiles, attributed to reduced carrier mobility and increased impurity scattering at higher $T$. Among all the proposed dopants, the Y-doped $\alpha$-Fe$_2$O$_3$ demonstrated the highest $\sigma/\tau$, which can be linked to its lower $m^*$, higher $v_g$, and greater $n_{\rm eff}$. Additionally, the co-doping of (B, Y) and (B, Nb) resulted in $\sigma/\tau$ values comparable to the Y-doped $\alpha$-Fe$_2$O$_3$. 

At a higher $N$, the introduction of defects created additional trap sites and increased carrier scattering, necessitating more complex and tightly controlled synthesis methods. Therefore, we selected an optimal $N=10^{20}$ cm$^{-3}$, as changes in $\sigma/\tau$ became negligible with higher $N$ under our proposed doping conditions. The $\sigma/\tau$ values for Y, Nb, (Y, B), and (Y, Nb) dopants were calculated to be $8.725 \times 10^{18}$, $4.883 \times 10^{18}$, $7.870 \times 10^{18}$, and $8.636 \times 10^{18}$ Sm$^{-1}$s$^{-1}$, respectively, at $N = 10^{20}$ cm$^{-3}$ and $T = 300$ K. These values surpass those of pristine $\alpha$-Fe$_2$O$_3$ with a monovalent dopant, which is $4.733 \times 10^{17}$ Sm$^{-1}$s$^{-1}$, by more than tenfold. These enhancements indicate that the doped $\alpha$-Fe$_2$O$_3$ photoelectrodes can generate increased photoinduced current while minimizing recombination and resistive losses.  

%
\begin{figure}[htbp]
    \centering
    \includegraphics[width =1.0\linewidth]{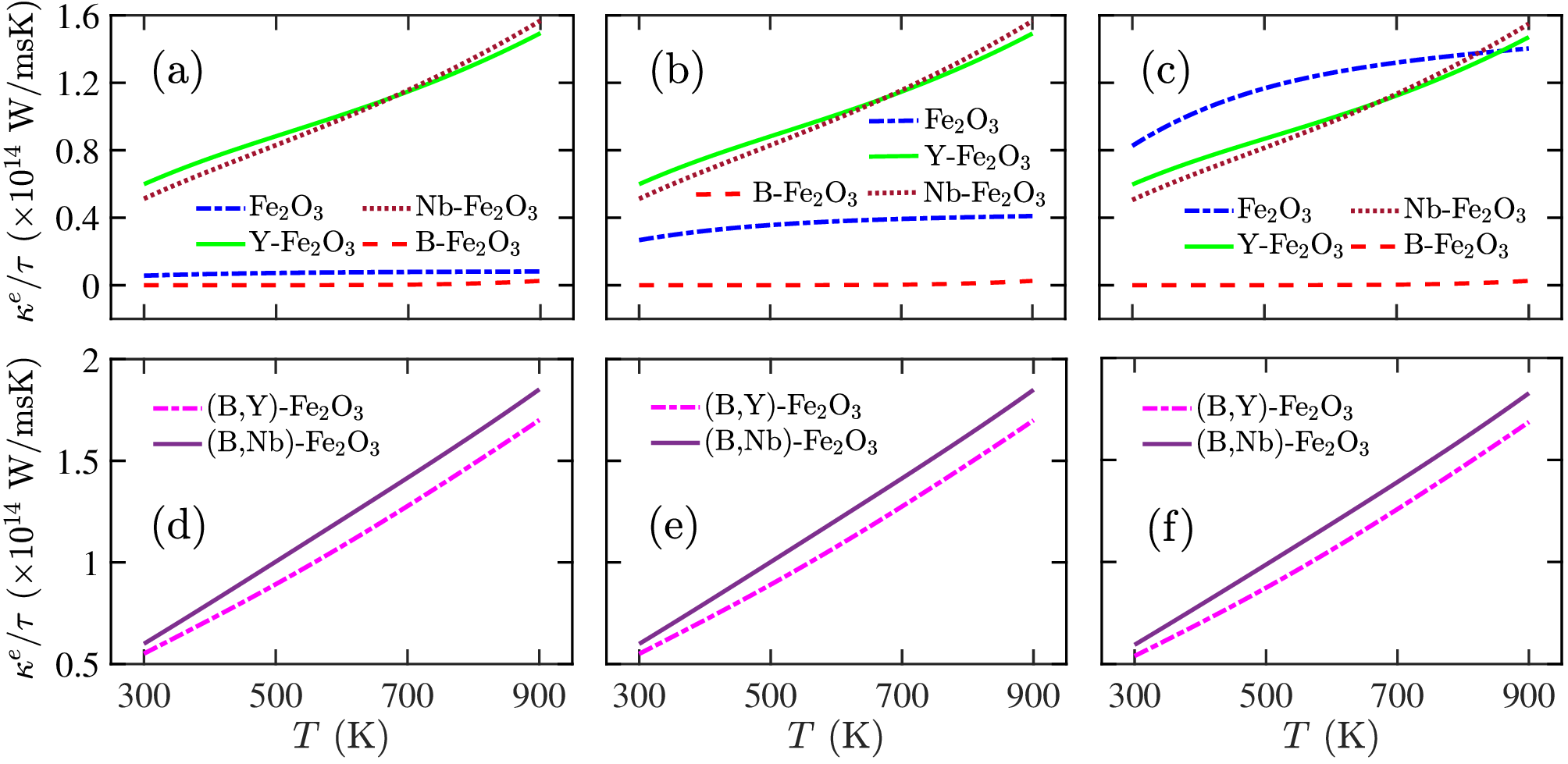}
    \caption{Electronic thermal conductivity divided by relaxation time ($\kappa^e$/$\tau$) as a function of temperature ($T$) for pristine pristine $\alpha$-Fe$_2$O$_3$, Y-doped $\alpha$-Fe$_2$O$_3$, Nb-doped $\alpha$-Fe$_2$O$_3$, and B-doped $\alpha$-Fe$_2$O$_3$ at various doping densities ($N$) of (a) $10^{19}$ cm$^{-3}$, (b) $10^{20}$ cm$^{-3}$, and (c) $10^{21}$ cm$^{-3}$. Electronic thermal conductivity divided by relaxation time ($\kappa^e$/$\tau$) as a function of temperature ($T$) for (B,Y)-doped $\alpha$-Fe$_2$O$_3$ and (B,Nb)-doped $\alpha$-Fe$_2$O$_3$ at various doping densities ($N$) of (a) $10^{19}$ cm$^{-3}$, (b) $10^{20}$ cm$^{-3}$, and (c) $10^{21}$ cm$^{-3}$.}
    \label{Fig5}
\end{figure}
%
%

The performance of photoelectrode materials is greatly affected by the processes of electronic carrier transport, which generate localized heating within the materials. This heating can lead to fractures and defect sites, compromising the integrity of the materials and accelerating photochemical corrosion in electrolyte environments \cite{kim2023thermal}. The thermal gradients that develop between the photoelectrode and the electrolyte create instability within the water electrolyzer cell, which negatively impacts the efficiency and sustainability of photocatalytic water splitting. To address these critical issues, it is essential to effectively dissipate heat and maintain a uniform thermal distribution within the materials during the water-splitting process. 

The Y and Nb dopants showed higher $\kappa^e/\tau$ values compared to $\alpha$-Fe$_2$O$_3$ at $N = 10^{19}$ and $10^{20}$ cm$^{-3}$, as shown in Fig.~\ref{Fig5}(a)--(b). In contrast, $\alpha$-Fe$_2$O$_3$ demonstrated an increased $\kappa^e/\tau$ value at $N = 10^{21}$ cm$^{-3}$ under lower $T$, which can be attributed to high ionized impurities and scattering, as presented in Fig.~\ref{Fig5}(c). For the co-dopants of (B, Y) and (B, Nb), the measured $\kappa^e/\tau$ values outperformed those of other doped photoelectrodes across a range of $N = 10^{19}$ to $10^{21}$ cm$^{-3}$, as illustrated in Fig.~\ref{Fig5}(d)--(f). The values of $\sigma/\tau$ increased with $T$ ranging from 300 to 500 K for all doped photoelectrodes, except for B mono-doping, as the $\sigma/\tau$ is directly related to $T$ and carrier concentration. At $N = 10^{20}$ cm$^{-3}$ and $T = 300$ K, the calculated $\kappa^e/\tau$ values for Y, Nb, (B, Y), and (B, Nb) dopants were $5.987 \times 10^{13}$, $5.122 \times 10^{13}$, $5.501 \times 10^{13}$, and $5.991 \times 10^{13}$ W/msK, respectively, each more than double the value for pristine $\alpha$-Fe$_2$O$_3$ with a monovalent dopant. The values of $C$  exhibited trends similar to those of $\kappa^e/\tau$ for pristine and doped $\alpha$-Fe$_2$O$_3$ photoelectrodes, as shown in Fig.~S2. These improvements enable efficient heat dissipation and maintain the stability of photoelectrodes during solar-driven water electrolysis operation. 

%
\begin{figure}[htbp]
    \centering
    \includegraphics[width =0.81\linewidth]{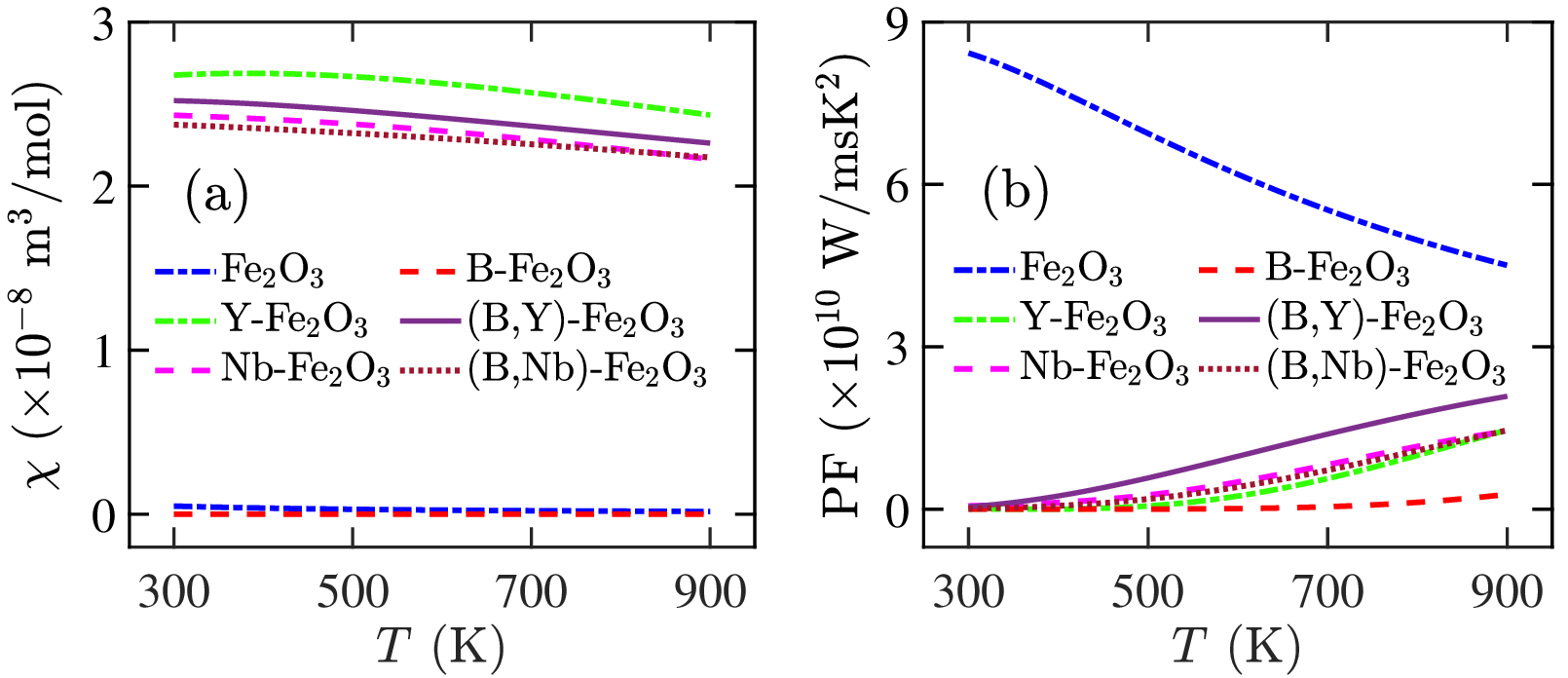}
    \caption{(a) Pauli magnetic susceptibility ($\chi$) and (b) power factor (PF) as a function of temperature ($T$) at doping density ($N$) of $10^{20}$ cm$^{-3}$ for pristine $\alpha$-Fe$_2$O$_3$, Y-doped $\alpha$-Fe$_2$O$_3$, Nb-doped $\alpha$-Fe$_2$O$_3$, B-doped $\alpha$-Fe$_2$O$_3$, (B, Y)-doped $\alpha$-Fe$_2$O$_3$, and (B, Nb)-doped $\alpha$-Fe$_2$O$_3$.}
    \label{Fig6}
\end{figure}
%
%

The parameter $\chi$ indicates the potential for generating spin-polarized currents, as well as the strength of the magnetic field, both of which drive efficient charge separation and enhance photoinduced current \cite{he2025optimizing}. Figure \ref{Fig6}(a) shows the values of $\chi$ for both pristine and doped $\alpha$-Fe$_2$O$_3$ photoelectrodes at $N = 10^{20}$ cm$^{-3}$ and $T$ ranging from $300$ to $900$ K. The Y, Nb, (B, Y), and (B, Nb) dopants exhibited an increase in $\chi$ by more than two orders of magnitude compared to pristine $\alpha$-Fe$_2$O$_3$, indicating significant potential for these doped photoelectrodes to generate spin-selective polarized current. Additionally, the enhanced magnetic strength can enhance carrier separation efficiency and accelerate carrier transport toward the active sites at the photoelectrode interfaces. As the $T$ increases, the values of $\chi$ demonstrate a slight decline due to carrier saturation near the Fermi level and the CB. 

The PF serves as a critical performance parameter in thermoelectric devices and has a weak correlation with photocatalytic activity. The PF reflects the balance between the $S$ and conductivity, with an optimal value significantly enhancing the efficiency of converting thermal energy into electricity. Figure \ref{Fig6}(b) illustrates the PF values for pristine and doped $\alpha$-Fe$_2$O$_3$ photoelectrodes at $N = 10^{20}$ cm$^{-3}$. Pristine $\alpha$-Fe$_2$O$_3$ exhibited the highest PF values across a range of $T$ from $300$ to $900$ K due to its larger $S$. Conversely, the doped $\alpha$-Fe$_2$O$_3$ photoelectrodes demonstrated an increasing trend in the PF with rising $T$. While pristine $\alpha$-Fe$_2$O$_3$ shows a maximum PF, its very low $n_{\rm eff}$ and $\sigma$ limit its practical efficiency. Table \ref{Table2} presents the summarized outcomes for both pristine and doped $\alpha$-Fe$_2$O$_3$ photoelectrodes at $N=10^{20}$ cm$^{-3}$ and $T=300$ K. Regarding all aspects of solar-driven water electrolysis, the doped $\alpha$-Fe$_2$O$_3$ photoelectrodes, excluding the B mono-doping, appears to be a promising option for efficient and effective photoelectrodes used in sustainable green H$_2$ generation. 

%
%
\begin{table}[htbp]
\centering
\caption{Summary of thermoelectric charge transport properties for pristine and doped $\alpha$-Fe$_2$O$_3$ photoelectrodes at a temperature ($T$) of 300 K and a doping density ($N$) of $10^{20}$ cm$^{-3}$.}
\resizebox{0.999\textwidth}{!}{%
\begin{tabular}{c c c c c c c}
\Xhline{3\arrayrulewidth}
    Name & $\sigma$/$\tau$ & $\kappa^e$/$\tau$ & $\chi$ & $C$ & ZT$_e$ & PF \\
      & (Sm$^{-1}$s$^{-1}$) & (W/msk) & (m$^3$/mol) & (J/molK) & -- & (W/msK$^2$) \\
    \Xhline{2\arrayrulewidth}
    pristine  $\alpha$-Fe$_2$O$_3$  & $4.733 \times 10^{17}$ & $2.737 \times 10^{13}$ & $5.005 \times 10^{-10}$ & $5.782$ & $9.463 \times 10^{-1}$ & $8.425 \times 10^{10}$ \\
    Y-doped $\alpha$-Fe$_2$O$_3$  & $8.725 \times 10^{18}$ & $5.987 \times 10^{13}$ & $2.676 \times 10^{-8}$ & $47.440$ &  $3.383 \times 10^{-4}$ & $6.751 \times 10^{7}$ \\
    Nb-doped $\alpha$-Fe$_2$O$_3$  & $4.883 \times 10^{18}$ & $5.122 \times 10^{13}$ & $2.432 \times 10^{-8}$ & $41.360$ & $3.413 \times 10^{-3}$ & $5.828 \times 10^{8}$ \\
    B-doped $\alpha$-Fe$_2$O$_3$  & $9.149 \times 10^{8}$ & $1.178 \times 10^{6}$ & $1.998 \times 10^{-18}$ & $6.1 \times 10^{-7}$ & $9.962 \times 10^{-1}$ & $3.911 \times 10^{3}$ \\
    (B, Y)-doped $\alpha$-Fe$_2$O$_3$  & $7.870 \times 10^{18}$ & $5.501 \times 10^{13}$ & $2.521 \times 10^{-8}$ & $42.910$ & $2.474 \times 10^{-3}$ & $4.536 \times 10^{8}$ \\
    (B, Nb)-doped $\alpha$-Fe$_2$O$_3$  & $8.636 \times 10^{18}$ & $5.991 \times 10^{13}$ & $2.374 \times 10^{-8}$ & $40.170$ & $5.242 \times 10^{-4}$ & $1.047 \times 10^{8}$ \\
    \Xhline{3\arrayrulewidth}
\end{tabular}}
\label{Table2}
\end{table}
%
%

%
\subsection{Underlying Fundamental Mechanism for the Enhancement}

The proposed doping conditions for pristine $\alpha$-Fe$_2$O$_3$ significantly enhance its photoabsorption, free carrier concentration, and conductivity through several fundamental physicochemical mechanisms. The improved photoabsorption mainly results from the reduction of $E_g$, which occurs due to quantum mechanical interactions between the dopant atoms and the host material \cite{walsh2008origins}. These interactions include the electron-exchange interaction energy, changes in $m^*$, and orbital mixing in the CB and VB through hybridization. Such phenomena create new energy states within the electronic structure, leading to a downshift in the CB and an upshift in the VB, ultimately decreasing $E_g$. 

The reduced $E_g$ for the (B, Y) and (B, Nb) dopants allows for the absorption of sunlight up to a wavelength of approximately 1000 nm, making these compounds promising photoelectrodes in terms of optical absorption and also capable of producing sufficient photoinduced voltage for photocatalytic water splitting. The introduction of additional electrons or holes from the dopants facilitates an increase in free carrier concentration. The Y and Nb dopants substitute Fe atoms in $\alpha$-Fe$_2$O$_3$ and donate electrons by ionizing to Y$^{3+}$ and Nb$^{5+}$, respectively. This ionization leads to a shift of the Fermi level toward the CB. The change in the Fermi level results in an increased carrier concentration, as described by the equation \cite{streetman2000solid}
\begin{equation}
    n = N_{\rm C} \hspace{0.5mm} {\rm exp} \left({-\frac{E_{\rm C} - E_{\rm F}}{k_{\rm B}T}} \right), 
\end{equation}
where $N_{\rm C}$ is the effective density of states in the CB, and $k_{\rm B}$ and $E_C$ are the Boltzmann constant and the energy level of CB, respectively. 

Additionally, the reduction of $E_g$ increases the intrinsic carrier concentration, as $E_g$ has a negative exponential correlation with the carrier concentration. The Nb-doped $\alpha$-Fe$_2$O$_3$ exhibited the highest carrier concentration due to the proximity of the Fermi level to the CB. Conversely, the co-dopants containing B act as acceptors, which causes a downshift of the Fermi level. This downshift results in a slight decrease in carrier concentration when compared to the mono-dopants of Y and Nb.     

Several factors, including carrier concentration, band dispersion, and $m^*$, affect the $\sigma$. In pristine $\alpha$-Fe$_2$O$_3$, the overlapping 3d-orbitals of Fe atoms lead to weak band dispersion in the CB, resulting in localized charge carriers, which subsequently reduces carrier concentration. In contrast, when dopants are introduced, their valence d-orbitals interact with the 3d orbitals of Fe, creating new states that enhance both band dispersion and the DOS. This interaction can lead to a higher concentration of delocalized charge carriers and a lower $m^*$. Considering these factors, $\sigma$ can be expressed as \cite{streetman2000solid}
\begin{equation}
    \sigma = \frac{ \tau q^2  n_{\rm eff}}{m^*}, 
\end{equation}

The Y-doped $\alpha$-Fe$_2$O$_3$ exhibited a reduced $m^*$ at the CBM and an increase in carrier concentration, both of which enhance $\sigma$. While the Nb, (B, Y), and (B, Nb) dopants resulted in $m^*_{\rm CBM}$ approximately twice that of pristine $\alpha$-Fe$_2$O$_3$, the remarkably increased carrier concentration mitigates this adverse effect, leading to a significant improvement in $\sigma$. In scenarios of low to moderate doping densities, the $\sigma$ consistently increases with higher doping levels. However, at high doping densities, ionized impurity scattering and potential carrier localization can occur, which may suppress carrier mobility, flatten the band dispersion, and ultimately lead to conductivity saturation. Consequently, the changes in $\sigma$ for doped $\alpha$-Fe$_2$O$_3$ photoelectrodes become limited as the doping density varies in this work.  

The $\kappa^e$ is influenced by several factors, including carrier mobility, heat transfer through electrons or holes, and energy distribution. In photoelectrode materials, the $\kappa^e$ is directly related to $\sigma$ and $T$, as expressed by the Wiedemann-Franz law \cite{franz1853ueber}
\begin{equation}
    \kappa^e =L(T)\sigma T = \frac{L(T) \tau q^2 n_{\rm eff} T}{m^*}, 
\end{equation}
where $L(T)$ is the Lorenz number. The value of $\kappa^e$ depends on both the carrier concentration and $m^*$, which are affected by doping methods and tailored band dispersion. As the doped $\alpha$-Fe$_2$O$_3$ photoelectrodes, except for B mono-doping, showed remarkably increased carrier concentration and $\sigma$, which led to improved $\kappa^e$. Similarly, the $\chi$ is correlated with carrier concentration based on the free-electron model at the Fermi level as $\chi \propto [n(E)]^{1/3}$, which leads to enhanced $\chi$ for the doped $\alpha$-Fe$_2$O$_3$ photoelectrodes. 

Among all the doping conditions proposed in this work, B-doped, (B, Y)-doped, and (B, Nb)-doped $\alpha$-Fe$_2$O$_3$ exhibited a significantly reduced $E_g$, thereby enabling broader solar energy harvesting across the visible and infrared regions. However, the limited carrier transport in B-doped $\alpha$-Fe$_2$O$_3$ constrains its effectiveness for photocatalytic water splitting. The Y and Nb mono-dopants and the (B, Y) and (B, Nb) co-dopants in $\alpha$-Fe$_2$O$_3$ showed nearly identical charge transport characteristics, which significantly outperformed each electronic property compared to pristine $\alpha$-Fe$_2$O$_3$. Nevertheless, both Y-doped and Nb-doped $\alpha$-Fe$_2$O$_3$ possessed high $E_g$ values, limiting their ability to absorb light in the visible region. Therefore, the proposed co-doping conditions represent the best options for achieving excellent photoabsorption and superior charge transport dynamics in the solar-driven water-splitting process, ultimately contributing to sustainable and efficient green H$_2$ generation in renewable energy applications.   

%
%

%
%
\section{Conclusion} 

In summary, we focused on identifying optimal doping conditions to improve the optoelectronic charge transport and thermal properties of $\alpha$-Fe$_2$O$_3$. We investigated B, Y, and Nb mono-dopants, as well as (B, Y) and (B, Nb) co-dopants. The use of B, (B, Y), and (B, Nb) significantly improved photoabsorption in the visible spectrum. While B mono-doping created impurity bands that hindered charge transport, the addition of Y and Nb dopants resulted in slightly reduced $E_g$ and substantial improvements in $n_{\rm eff}$, $\sigma/\tau$, $\kappa^e/\tau$, and $\chi$ compared to pristine $\alpha$-Fe$_2$O$_3$. 

The co-doping strategies (B, Y) and (B, Nb) demonstrated remarkable enhancements in electronic charge transport dynamics, along with exceptional photoabsorption. Considering all aspects of photoelectrochemical water splitting, the co-doping combinations of (B, Y) and (B, Nb) emerged as the best choices for solar energy harvesting, carrier transport characteristics, and thermal performance, highlighting their potential as high-performance photoelectrodes for sustainable green H$_2$ generation through solar-driven water electrolysis. In addition to their applications in photoelectrochemistry, these photoelectrode materials can also be utilized in thermoelectric systems to convert thermal energy into electricity efficiently. They exhibit overall improved thermoelectric performance compared to pristine $\alpha$-Fe$_2$O$_3$, demonstrating their versatility and potential for various renewable energy solutions.         

%
%

\section*{Supplementary Material}
The supplementary material presents the optimized lattice parameters and band gap energies for pristine and doped $\alpha$-Fe$_2$O$_3$. The supplementary material also displays the temperature dependence of the Seebeck coefficient and specific heat for various doping densities.  

\section*{Data Availability}
All data of the paper are presented in the main text and the supplementary material.

%
\section*{Author Declaration} 
The authors have no conflicts to disclose.

%

%
%
\small
\bibliographystyle{ieeetr}
\bibliography{references}

%
\end{document}